\documentclass[twocolumn,showpacs,preprintnumbers,amsmath,amssymb]{revtex4}

\usepackage[dvips]{hyperref}
\usepackage{graphicx}
\usepackage{dcolumn}
\usepackage{bm}

    \newcommand{\crs}{CeRu$_{2}$Si$_2$~}
    \newcommand{\crg}{CeRu$_{2}$Ge$_2$~}
    \newcommand{\lrg}{LaRu$_{2}$Ge$_2$~}
    \newcommand{\ccs}{CeCu$_{2}$Si$_2$~}
    \newcommand{\ccg}{CeCu$_{2}$Ge$_2$~}
    \newcommand{\cps}{CePd$_{2}$Si$_2$~}
    \newcommand{\cpg}{CePd$_{2}$Ge$_2$~}

\begin{document}

\title{Probing the phase diagram of CeRu$_2$Ge$_2$ by thermopower at high pressure}

\author{Heribert Wilhelm}

\affiliation{Max--Planck--Institut f\"ur Chemische Physik fester Stoffe,
  N\"othnitzer Str. 40, 01187 Dresden, Germany}

\author{Didier Jaccard}

\affiliation{D\'epartement de Physique de la Mati\`ere Condens\'ee,
  Universit\'e de Gen\`eve, Quai Ernest-Ansermet 24, 1211 Gen\`eve 4,
  Switzerland}



\begin{abstract}
The temperature dependence of the thermoelectric power, $S(T)$, and
the electrical resistivity of the magnetically ordered \crg
($T_N=8.55$~K and $T_C=7.40$~K) were measured for pressures $p <
16$~GPa in the temperature range 1.2~K~$<T<300$~K.  Long-range
magnetic order is suppressed at $p_c\approx 6.4$~GPa. Pressure drives $S(T)$
through a sequence of temperature dependences, ranging from a
behaviour characteristic for magnetically ordered heavy fermion
compounds to a typical behaviour of intermediate-valent systems. At
intermediate pressures a large positive maximum develops above 10~K in
$S(T)$. Its origin is attributed to the Kondo effect and its position
is assumed to reflect the Kondo temperature $T_K$. The pressure
dependence of $T_K$ is discussed in a revised and extended $(T,p)$
phase diagram of CeRu$_2$Ge$_2$.
\end{abstract}


\maketitle

\section{Introduction}
\label{sec:introduction} 
A large number of heavy fermion (HF) or unstable valence compounds and
alloys revealed a complicated temperature ($T$) dependence of the
thermoelectric power ($S$). This is due to competing interactions
present in these systems and the sensitivity of $S$ to details of the
band structure. Depending on the strength of the hybridisation between
4f and conduction electrons, the exchange interaction $J$ leads to
long-range magnetic order, magnetic Kondo systems, HF or
intermediate-valence (IV) behaviour. As a measure for the strength of
$J$ of these four regimes one can use the Kondo temperature $T_K$ and
the N\'eel temperature $T_N$ (or the Curie temperature $T_C$). Sakurai
and coworkers \cite{Sakurai96} used the ratio of $T_K$ to $T_N$ to
classify the $S(T)$ data of different compounds into these regimes,
each showing a characteristic $S(T)$ dependence.  Ce-based
representatives of each regime investigated by $S(T)$ are
CeAu$_2$Si$_2$ \cite{Amato85}, CeAl$_2$ \cite{Jaccard82}, CeAl$_3$
\cite{Jaccard85a} and \crs \cite{Amato88,Amato89} or CeNi$_2$Si$_2$
\cite{Sampathkumaran89,Levin81}.

Typical features in $S(T)$ of these regimes can be seen in a single
compound if $J$ is increased, e.~g.~by alloying. Increasing $x$ in the
pseudo-binary alloy Ce(Pb$_{1-x}$Sn$_x$)$_3$ \cite{Sakurai88} and
in the solid-solution CeRh$_{2-x}$Ni$_{x}$Si$_2$
\cite{Sampathkumaran89} tunes the systems from trivalent to
intermediate-valent, i.e.~enhancing $J$. Also the interatomic
distances seem to control the shape of $S(T)$. This results from an
analysis of the $S(T)$ data of many Ce$M_2X_2$ compounds, with $M$ a
transition metal and $X=$~Si or Ge \cite{Jaccard92}. Apparently, in
the series Ce$M_2$Si$_2$, where the corrected interatomic distances
decrease for $M=$~Au, Cu, Rh, Ru, and Fe, the increase of $J$ is
accompanied with a characteristic variation of $S(T)$ at low
temperatures.

This systematic volume dependence suggests that $S(T)$ is very
sensitive to pressure. Based on the pressure-induced changes in
$S(T)$, Link et al.~\cite{Link96} sketched a sequence of $S(T)$
dependences that shows the evolution of characteristic features in
$S(T)$ with pressure. The $S(T)$ of a magnetically ordered system,
like \ccg shows one negative and one positive peak below $\approx
20$~K and above about 100~K, respectively \cite{Jaccard92}. In
general, pressure will suppress the negative peak and leads to the
appearance of new features at low temperatures. As pressure has tuned
the system into the IV regime, only one maximum well above room
temperature remains, like in CeNi$_2$Si$_2$ at ambient pressure
\cite{Levin81}. Similarly, the application of pressure on the
non-magnetic HF compounds \ccs ($T_K\approx 20$~K) \cite{Jaccard85}
and CeAl$_3$ ($T_K\approx 5$~K) \cite{Fierz88} reduces the magnitude
of the negative peak and leads to the development of two additional
positive peaks around 20~K and even lower temperature. The analysis of
these pressure-induced $S(T)$ dependences has led to the conjecture
that the pressure-induced maximum in $S(T)$ located at approximately
20~K is related to the Kondo effect \cite{Link96}. The origin of the
high-temperature maximum, present already at low pressure, is due to
the crystal-field separation of the Ce 4f electron energy level from
the ground state \cite{Bhattacharjee76}. Zlati\'{c} and coworkers
\cite{Zlatic2003} achieved a qualitative understanding of the
experimental results, using the
Coqblin-Schrieffer model and assuming a splitting of the 4f
states in the presence of a crystalline electric field.

The magnetically ordered \crg offers the possibility to map the
pressure dependence of the characteristic features in $S(T)$ since its
pressure-induced transition into the HF regime was intensively studied
\cite{Wilhelm98,Wilhelm99,Kobayashi98,Suellow99,Bouquet2000,Demuer2000}.
The ambient pressure $S(T)$ curve \cite{Wilhelm99} contains no maximum
below room temperature, which is a consequence of the large energy
separation of the crystal-field levels ($\Delta_1=500$~K and
$\Delta_2=750$~K \cite{Felten87,Loidl92}). Assuming that the crystal
field levels are not strongly influenced by pressure, two distinct
positive peaks at low and high temperature in $S(T)$ at intermediate
pressures are expected. This could provide insight into the pressure
dependence of $T_K$. Moreover, the $S(T)$ data might reveal
information about the pressure-induced change from a magnetically
ordered compound (at $p=0$) to a HF Kondo-lattice compound (at high
pressure), equivalent to \crs ($T_K=24$~K \cite{Besnus85}) at ambient
pressure.


\section{Experimental Details}
\label{sec:experimentaldetails} The \crg and \lrg samples have been prepared
by arc-melting stoichiometric amounts of Ce, La, Ru, and Ge under
argon atmosphere. The purity of the elements was 99.99\%, except for
Ge, which had a purity of 99.999\%. X-ray diffraction pattern could be
indexed according to the ThCr$_2$Si$_2$ structure ($I4/mmm$) with
lattice parameters $a=4.2685(4)$~\AA~ and $c=10.048(3)$~\AA~ for
CeRu$_2$Ge$_2$ and $a=4.314(2)$~\AA~ and $c=10.129(6)$~\AA~ for
LaRu$_2$Ge$_2$. Samples of the \crg ingot have been used in earlier
electrical resistivity, $\rho(T)$ \cite{Wilhelm98,Wilhelm99}, and
specific heat \cite{Bouquet2000} pressure studies.

\begin{figure}
\center{\includegraphics[width=0.80\columnwidth,clip]{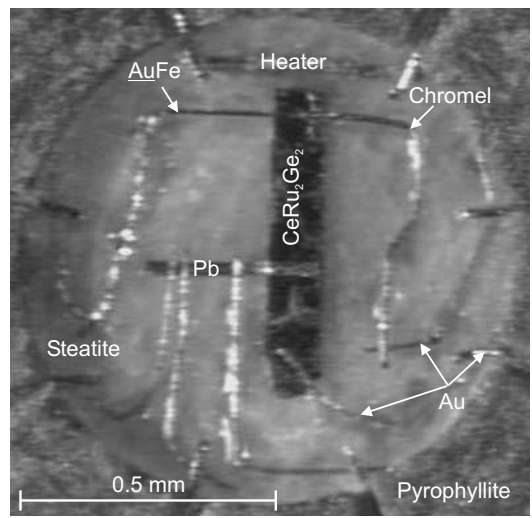}}
\caption{Top view of the inner part of the pressure chamber before
closing. Close to the heater two thermocouples (\underline{Au}Fe
and Chromel) are located on top of the sample. The Au-wire
connected at the opposite edge of the sample is chosen as
reference point for the two thermovoltage $V_{\rm{\underline
Au}Fe}$ and $V_{\rm Chromel}$. Lead is used as pressure gauge.
Au-wires establish the connection through the pyrophyllite
gasket. Steatite serves as pressure transmitting medium.} \label{fig:cell}
\end{figure}

A clamped Bridgman anvil cell with synthetic diamonds was used to
measure $S(T)$ and $\rho(T)$ in the temperature range $1.2~{\rm K} < T
< 300~{\rm K}$. The pressure chamber was made of a non-metallic gasket
(pyrophyllite, $\phi_{\rm int} = 1$~mm) and two steatite disks served
as pressure transmitting medium. Electrical leads were attached to the
sample (cross-section of $14\times 108\mu$m$^2$) in such a way that a
four-point $\rho(T)$ and, in a separate run, a $S(T)$ measurement
could be performed (Fig.~\ref{fig:cell}). The pressure dependence of
the superconducting transition temperature of Pb yielded the pressure
\cite{Bireckhoven88}.

A heater (Chromel wire), located close to the small edge of the
sample, produced a temperature gradient $\Delta T$ along the
sample. The opposite edge of the sample remained at $T_0$, the
temperature of the pressure cell \cite{Jaccard98}, and served as
reference for the two thermocouples \underline{Au}Fe (with 0.07 at\%
Fe) and Chromel. The two measured thermovoltages were
$V_{\rm\underline{Au}Fe} =(S_{\rm \underline{Au}Fe}-S)\Delta T$ and
$V_{\rm Chromel}=(S_{\rm Chromel}-S)\Delta T$. The absolute
thermoelectric power of the sample, $S$, at $T_0 + \Delta T/2$ is
given by:
\begin{equation}
S = S_{\rm\underline{Au}Fe} +
\frac{S_{\rm{Chromel}}-S_{\rm\underline{Au}Fe}}{1-V_{\rm
Chromel}/V_{\rm\underline{Au}Fe}}~.
\end{equation}
\noindent 

The absolute thermopower of \underline{Au}Fe and Chromel are assumed to be
pressure independent. This seems to be a good assumption since the absolute
value of $S_{\rm\underline{Au}Fe}$ at 12~GPa and 4.2~K is only 20\% smaller
than at ambient pressure \cite{Wilhelm2002}. However, small pressure-induced
changes in $S(T)$ of the sample should be interpreted carefully.

\section{Results}
\label{results}
\begin{figure}
\center{\includegraphics[width=1.0\columnwidth,clip]{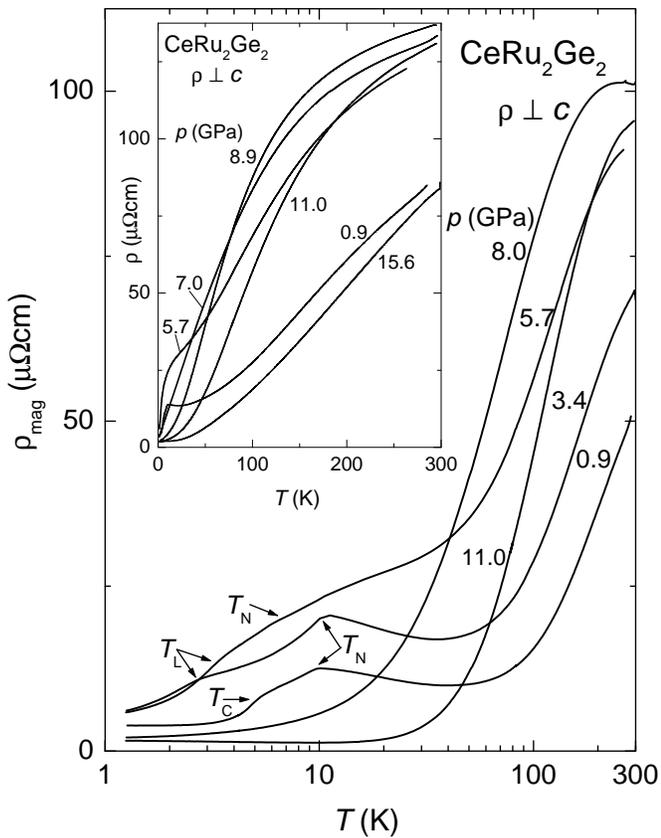}}
\caption{Magnetic contribution $\rho_{mag}(T)$ to the electrical
  resistivity of \crg at different pressures. Two different
  antiferromagnetic phases occur below $T_N$ and $T_L$. A ferromagnetic
  ground state is present below $T_C$ and low pressure. No traces
  of magnetic order are observed for $p>7$~GPa above 1.2~K. Inset: Raw
  data of $\rho(T)$ in a linear plot. Incoherent Kondo scattering is
  apparent for $p=5.7$~GPa.}
\label{fig:crgrhotep}
\end{figure}

Figure~\ref{fig:crgrhotep} shows the magnetic part, $\rho_{mag}(T)$,
of $\rho(T)$ of \crg at several pressures. It was obtained by
subtracting a phonon contribution, approximated as
$\rho_{ph}(T)=0.12~\mu\Omega$cm/K$\times T$, from the raw data shown
in the inset to Fig.~\ref{fig:crgrhotep}. The slope
$\partial\rho(T)/\partial T = 0.12~\mu\Omega$cm/K was deduced from our
$\rho(T)$ measurement of the non-magnetic reference compound \lrg at
ambient pressure for $T>70$~K. This approximation had to be used since
the measured value of $\rho(T)$ of \lrg is slightly overestimated,
presumably due to microcracks. At low temperature $\rho_{mag}(T)$ is
dominated by the magnetic phase transitions, manifested by several
discontinuities in $\rho_{mag}(T)$. The transition temperatures were
defined by the intersection of two tangents drawn to the $\rho(T)$
curve below and above the kink. A high-temperature peak below room
temperature evolves for pressures in the range $7.0~\leq p\leq
10.4$~GPa. It is due to the interplay of the Kondo and crystal-field
effects. At intermediate temperatures and pressures (e.g. at about
20~K and 5.7~GPa in inset to Fig.~\ref{fig:crgrhotep}), incoherent
Kondo scattering is clearly increasing. In contrast to other compounds
like \cpg \cite{Wilhelm2002}, CeCu$_2$Ge$_2$ \cite{Jaccard99}, and
CeCu$_5$Au \cite{Wilhelm2000} its contribution cannot be deconvoluted
due to its modest magnitude.

%
%

\begin{figure}
\center{\includegraphics[width=1.0\columnwidth,clip]{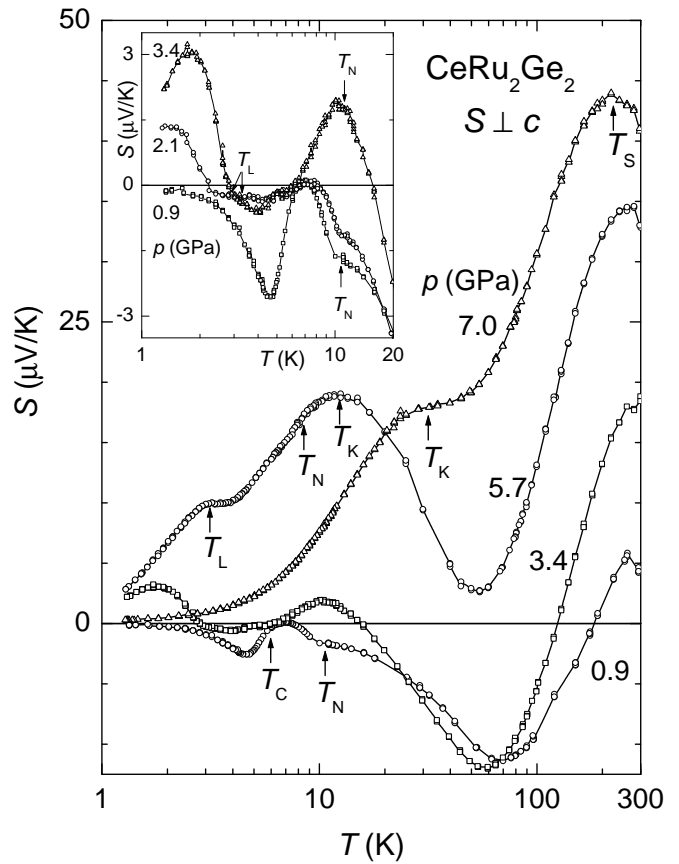}}
\caption{Temperature dependence of the thermoelectric power $S(T)$ of
\crg at selected pressures. The entrance into the magnetically ordered
states is indicated by $T_N$, $T_C$, and $T_L$, as deduced from the
$\rho(T)$ data. $T_K$ and $T_S$ label the centre of broad,
pressure-induced maxima. The inset shows the low-temperature
part of $S(T)$ at low pressure.}
\label{fig:teplowp}
\end{figure}

\begin{figure}
\center{\includegraphics[width=1.0\columnwidth,clip]{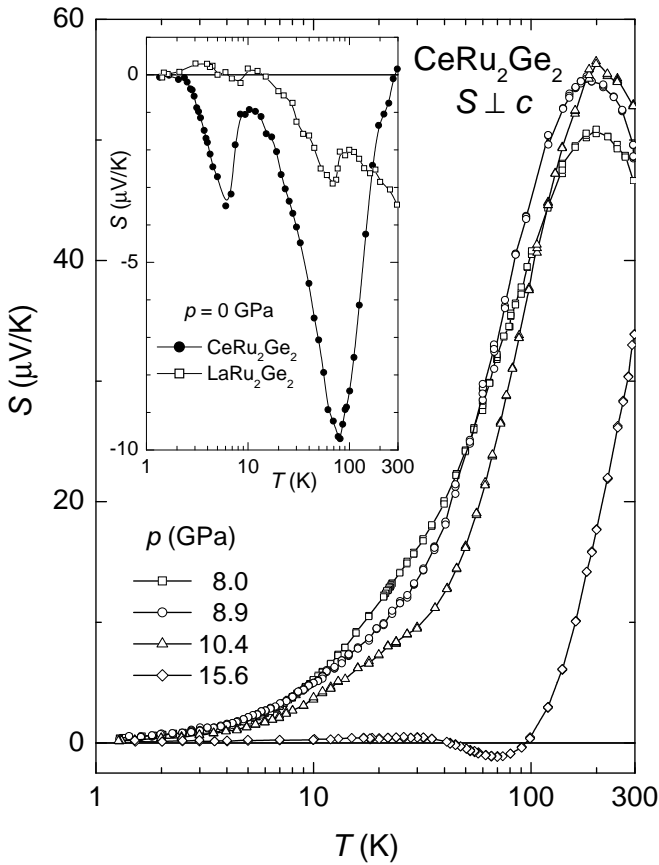}}
\caption{Thermoelectric power $S(T)$ of \crg above the critical
  pressure. $S(T)$ is dominated by a high-temperature
  maximum. The inset shows the ambient pressure $S(T)$ data of
  \crg \cite{Wilhelm99} and its non-magnetic counterpart LaRu$_2$Ge$_2$.}
\label{fig:tephighp}
\end{figure}

The interpretation of the high-temperature maximum in $\rho_{\rm
mag}(T)$ as a result of Kondo exchange interaction between the
conduction electrons and the crystal-field split ground state of the
Ce$^{3+}$-ions is supported by the evolution of the positive peak in
$S(T)$ below 300~K for $p>3.4$~GPa (see Figs.~\ref{fig:teplowp} and
\ref{fig:tephighp}). Its position $T_S$ corresponds to a fraction of
the crystal-field splitting as in many other Ce-based compounds and
alloys. In CeRu$_2$Ge$_2$, $T_S$ first decreases linearly with
pressure (-23~K/GPa), attains a minimum around 200~K at about 9~GPa,
and then starts to increase. Based on this $T_S(p)$ variation a
maximum in $S(T)$ is expected to occur at about 384~K at ambient
pressure, well above the limit of our set-up. The amplitude of the
peak grows linearly with pressure (6.4~$\mu$V/(K GPa)) and attains a
maximum value of 55~$\mu$V/K at about 10~GPa
(Fig.~\ref{fig:tephighp}).

%
%

%
%

The complicated $S(T)$ below 10~K (inset to Fig.~\ref{fig:teplowp}) is
very likely caused by the onset of magnetic order and to the opening
of a spin gap ($\Delta=14$~K
\cite{Wilhelm99,Raymond99}). Pressure-induced changes of the Fermi
surface due to the periodicity of the magnetic ordering (magnetic
super-zone effects) might be the explanation for the different
low-temperature $S(T)$ dependences for $p<3.4$~GPa.  At 5.7~GPa,
$S(T)$ is positive over the entire temperature range and a Kondo
maximum at about $T_K=12$~K (incoherent scattering on the ground
state) occurred (Fig.~\ref{fig:teplowp}). The position of the maximum
has shifted considerably towards higher temperatures in $S(T)$
recorded at 7.0~GPa, supporting its assignment to the Kondo effect. A
trace of this maximum can even be anticipated around 40~K at 8.0~GPa
as a weak shoulder at the high-temperature maximum entered at
$T_S\approx 200$~K (Fig.~\ref{fig:tephighp}). At sufficiently high
pressure both peaks merge. Then the Kondo maximum dominates the
high-temperature maximum as the crystal-field effect disappears when
the system enters the IV regime. At 15.6 GPa, the $S(T)$ maximum is
well above room temperature. Thus, the overall influence of pressure
on $S(T)$ of \crg fits rather well the behaviour sketched in
Ref.~\cite{Link96}.

The effect of magnetic ordering in $S(T)$ is obvious as \crg is
compared with the non-magnetic LaRu$_2$Ge$_2$ (inset to
Fig.~\ref{fig:tephighp}). \lrg has a rather small $S$ ($-3.5~\mu$V/K
at 300~K) which decreases almost linearly in magnitude with decreasing
temperature and reaches $S\approx 0$ at $T\approx10$~K
\cite{oscillations,Bando93}. Thus, the low-temperature anomalies in
$S(T)$ of \crg at ambient pressure are caused by the occurrence of
magnetic order. This implies that $S(T)$ of \crg above $T_N$ can be
regarded as the superposition of the pressure independent linear-in
temperature electron-phonon term represented by LaRu$_2$Ge$_2$ and the
incipient contribution characteristic of several HF compounds. That
contribution has a negative peak centred at 80~K (ascribed to spin
interactions \cite{Link96}) and a positive peak above room temperature
(caused by the interplay of Kondo and crystal-field effects). At
ambient pressure, the absolute value of $S$ is small (as for
e.~g.~CeAu$_2$Si$_2$ \cite{Link97}) because the compound is far below
the critical pressure.


\section{Discussion}
\label{discussion}

%
%
\begin{figure} 
\center{\includegraphics[width=1.0\columnwidth,clip]{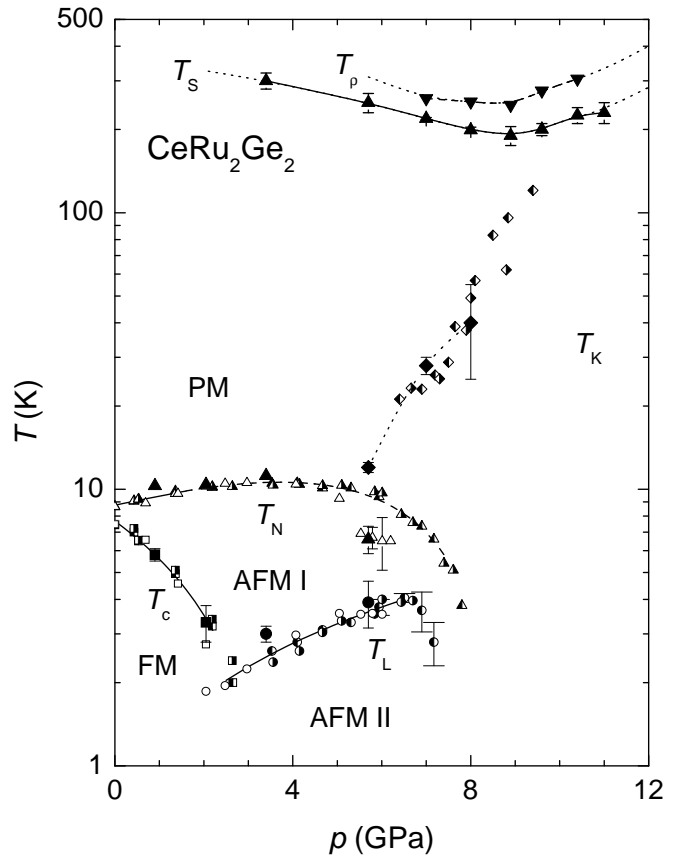}}
\caption{($T,p$) phase diagram of \crg obtained from electrical
  resistivity (half open symbols) \cite{Wilhelm98,Wilhelm99},
  calorimetric (open symbols) \cite{Bouquet2000}, and the combined
  $\rho(T)$ and $S(T)$ (bold symbols) measurements. At low pressure a
  paramagnetic (PM) to antiferromagnetic (AFM I) phase transition
  occurs at $T_N$ and a subsequent transition into a ferromagnetic
  phase (FM) takes place at $T_C$. The FM ground state is suppressed
  at 2.3(2)~GPa and a second antiferromagnetic\-ally ordered phase
  (AFM II) occurs below $T_L$. The combination of {\it all} data suggests
  that long-range magnetic order is suppressed at a critical pressure
  $p_c=7.8$~GPa. $T_{\rho}$ is the position of a maximum in $\rho(T)$
  whereas $T_S$ and $T_K$ represent the centre of peaks in $S(T)$. The
  half filled diamonds indicate $T_K \propto 1/\sqrt{A}$
  \cite{Wilhelm99}. They are shifted by 1.4~GPa towards lower pressure
  (see text).}
\label{fig:phasediagram}
\end{figure}

Figure~\ref{fig:phasediagram} shows a revised and extended ($T,p$)
phase diagram of \crg based on the present $S(T)$ and $\rho(T)$ data
as well as $\rho(T)$ and calorimetric experiments performed on samples
of the same batch \cite{Wilhelm98,Wilhelm99,Bouquet2000}. The
different magnetic phases in \crg are assumed to be the same as those
found in the solid-solution CeRu$_2$(Si$_{1-x}$Ge$_x$)$_2$ for $0\leq
x \leq 1$ \cite{Haen96,Haen2002}. For details about the magnetic
structures the reader is referred to Ref.~\cite{Raymond99} and
references therein.  Pressure on \crg or replacing Ge by Si in
CeRu$_2$(Si$_{1-x}$Ge$_x$)$_2$ are equivalent, since the unit-cell
volume seems to be the crucial parameter to change $J$. This was
concluded from the common $(T,V)$ phase diagram of both systems
\cite{Wilhelm99,Suellow99}. The main observation of interest is that
the long-range magnetic order is suppressed above a critical pressure
$p_c$ and a HF behaviour equivalent to \crs at ambient pressure is
expected. The present data show that the long-range magnetic order was
suppressed between 5.7~GPa and 7~GPa. Therefore, we will assume
$p_c\approx 6.4$~GPa as a rough estimate. Using all $p_c$ values
determined on samples of the same batch
\cite{Wilhelm98,Wilhelm99,Bouquet2000} including this work, we
estimate $p_c=7.8$~GPa for CeRu$_2$Ge$_2$. Depending on the
experimental method a slightly lower ($p_c \approx 6.9$~GPa
\cite{Bouquet2000}) or a higher value ($p_c=8.7$~GPa) of $p_c$ was
reported in Ref.~\cite{Wilhelm98,Wilhelm99}. This is partly due to the
criterion used for the determination of $p_c$. As far as $\rho(T)$
data of Ref.~\cite{Wilhelm99} are concerned, the pressures where the
anomalies in the $\tilde{A}(p)$ (at $p=7.8$~GPa) or $n(p)$ (at
$p=8.2$~GPa) dependences occur are a better choice of $p_c$ than the
$T_N(p)\rightarrow 0$ extrapolation ($p=8.7$~GPa). $\tilde{A}$ and $n$
were adjustable parameters in a fit of $\rho(T)=\rho_0 + \tilde{A}T^n$
to the data below 1.5~K, with $\rho_0$ the measured residual
resistivity.

Knowing the ($T,p$) phase diagram one is tempted to label the features
in $S(T)$ (Fig.~\ref{fig:teplowp}). It seems that a discontinuity in
the slope of $S(T)$ can be used as definition of $T_N$ for $p \leq
2.1$~GPa, similar to the case of \ccg \cite{Jaccard92}. At $p\geq
5.7$~GPa however, $T_N$ yields no signature in $S(T)$ as e.~g.~in the
case of CeAl$_2$ ($T_N = 3.8$~K \cite{Jaccard82} at $p=0$). From the
$\rho(T)$ data recorded at 5.7~GPa it is inferred that $T_N$ is
located slightly below the maximum of the broad peak centred at
12~K. The signature of the FM transition is obscure. If it is related
to the strong decrease of $S(T)$ below about 7~K (for $p \leq
0.9$~GPa) then the rather constant $S(T)$ between 2~K and 5~K found at
2.1~GPa (inset to Fig.~\ref{fig:teplowp}) would indicate that the
ferromagnetic order is already suppressed. The maximum at about 3~K in
the 5.7~GPa data seems to be caused by $T_L$. The assignment of the
remaining, strongly pressure dependent pronounced maximum as $T_K$ is
then quite obvious for $p\geq 5.7$~GPa. It might be speculated whether
the broad feature below 2~K (for $p\leq 3.4$~GPa) is a signature
of $T_K$ or related to the magnetic order.

%
%
\begin{figure}
\center{\includegraphics[width=85mm,clip]{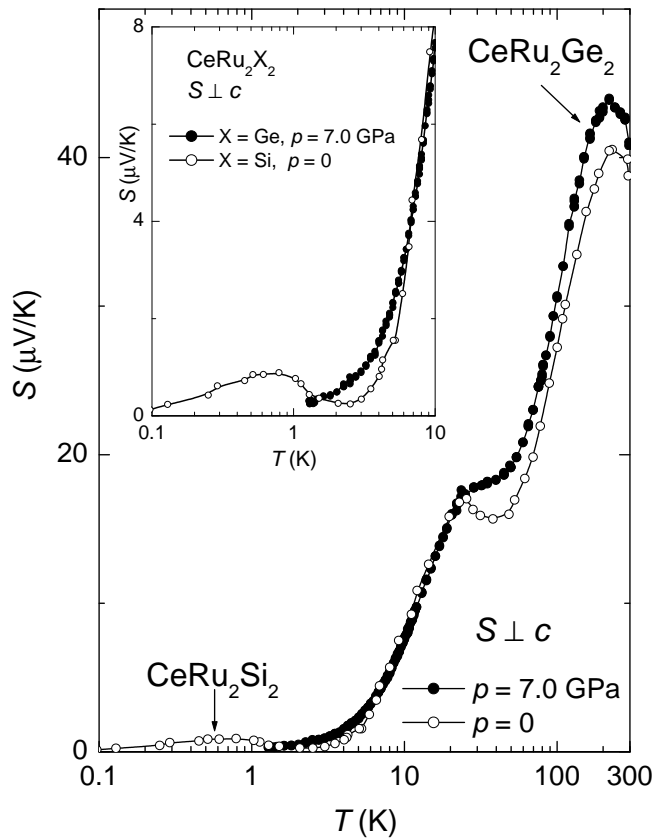}}
\caption{Comparison of $S(T)$ of \crg at 7.0~GPa and \crs at ambient
  pressure \cite{Amato88a}. Both curves reveal a pronounced
  high-temperature maximum and a moderate positive contribution to
  $S(T)$ at about 25~K. Inset: The low temperature part of $S(T)$ of \crs
  contains an additional positive peak below 1~K.}
\label{fig:crgcrstep}
\end{figure}

The interpretation of this pronounced low-temperature feature as
signature of the Kondo effect is supported by the comparison of $S(T)$
of \crg at 7.0~GPa and \crs at ambient pressure \cite{Amato89}
(Fig.~\ref{fig:crgcrstep}). It is important to note that in both cases
$S(T)$ was measured perpendicular to the c-axis of the tetragonal
crystal structure. The low-temperature maximum in \crs occurs at 25~K
close to the value of $T_K=24$~K extracted from specific heat
measurements using a single-impurity model
\cite{Besnus85}. Reminiscent to this is the weak maximum in $S(T)$ at
28~K for \crg at 7.0~GPa. This agreement is not accidental since also
the high-temperature maximum occurs at the same temperature $T_S$ and
has a similar magnitude as for CeRu$_2$Si$_2$. In $S(T)$ of \crs it
coincides with the temperature of the first excited crystal-field
level ($\Delta_1=220$~K), deduced from a Schottky anomaly in the
specific heat \cite{Besnus85}. Such a perfect agreement of
calorimetric and transport data is fortuitous since it depends on
$T_K$, the crystal-field splitting, and the degeneracy of the
crystal-field levels \cite{Maekawa85}. However, it applies also to
CeCu$_6$ \cite{Amato88a} and CePd$_2$Si$_2$ \cite{Link96}. Thus, the
peak in $S(T)$ of \crg at $T_S$ can be ascribed to the interplay
between incoherent Kondo scattering and crystal-field effects. This
comparison shows that the first excited crystal-field level in \crg
decreased from about 500~K \cite{Felten87,Loidl92} to about 200~K for
pressures in the vicinity of $p_c$ and explains why a high-temperature
maximum in $\rho(T)$ (Fig.~\ref{fig:crgrhotep}) was only seen in a
certain pressure range above $p_c$. Furthermore, the resemblance of
both $S(T)$ curves confirms the assumption of $p_c\approx 6.4$~GPa.

The qualitative and quantitative agreement of both data sets can even
be used to state that the occurrence of an inflection point in $S(T)$
of \crg at about 50~K might imply the change of regime as in \crs
\cite{Amato89}. Slightly above this temperature short-range AFM
inter-site correlations emerge in \crs at ambient pressure
\cite{Regnault88,Rossat88}. In addition a $\mu$SR investigation
\cite{Amato94} revealed magnetic correlations below 1~K. These
quasi-static correlations involve very small magnetic moments of the
order of $10^{-3}\mu_B$. The origin of this weak static magnetism well
below $T_K$ might be due to the formation of HF magnetism involving
the renormalised quasi-particles \cite{Amato94}. The presence of such
correlations might be responsible for the positive contribution to
$S(T)$ below 1~K (inset to Fig.~\ref{fig:crgcrstep}) and can be
expected to occur also in \crg below 1~K.

The very rapid increase of $T_K$ with pressure is corroborated by the
reported pressure dependence of the $A$-coefficient \cite{Wilhelm99},
which was deduced from a fit of $\rho(T)=\rho_0 + AT^2$ to the data
for $T<0.5$~K. Assuming that the relation $A\propto 1/T_K^2$ holds, we
calculated $T_K(p)$ and added it to the phase diagram
(Fig.~\ref{fig:phasediagram}). In order to do so, these values were
normalised in such a way that at 7.0~GPa $T_K=24$~K was in agreement
with $T_K$ deduced from the $S(T)$ data. Furthermore, the calculated
$T_K(p)$ data had to be shifted towards lower pressure ($\Delta p_c=-1.4$~GPa)
to account for the different $p_c$ values. The normalisation is in part
arbitrary because the $S(T)$ contribution gives only an estimation of
$T_K$. But the pressure variations are reliable and noteworthy. Note
here that one compares the energy scale $T_K$ with the ground state
excitations, implying that only one energy scale exists and that
notably $T_K$ is finite at $p_c$. The $T_K(p)$ dependence revealed the
interesting finding that two pressure ranges exist with quite
different slopes in the $T_K$ vs. $p$ plot. Comparable $\partial
T_K/\partial p$ values are observed for pressures ranging from about
6.5~GPa to 7~GPa and above 8.5~GPa. A much larger slope is found at
intermediate pressures (7~GPa~$\leq p\leq 8.5$~GPa), i.e. around
$p_c$. The transition from a large to a small $\partial T_K/\partial
p$ value is reminiscent to the change in the slope of $T_K$ vs. $p$
observed in \crs at about 1~GPa \cite{Payer93}.  A weaker than
exponential variation of $T_K(p)$ due to intersite interaction,
proposed in Ref.~\cite{Iglesias97}, is not seen. If such a
trend exist it seems to be restricted to pressures well below $p_c$.

The extrapolation of $T_K(p)$ seems to merge with $T_S(p)$ at
pressures above 11~GPa, indicating the entrance into the IV regime
(Fig.~\ref{fig:phasediagram}). Here, $k_BT_K$ exceeds the
crystal-field splitting and the entire six-fold degeneracy of the
Ce-4f$^1$ multiplet is recovered. So far, similar pressure dependences
of two maxima in $\rho(T)$ were found for CeCu$_2$Ge$_2$
\cite{Jaccard99}, \cps \cite{Demuer2002}, \cpg \cite{Wilhelm2002}, and
CeCu$_5$Au \cite{Wilhelm2000}. The pressure $p_v\approx 11$~GPa for
CeRu$_2$Ge$_2$, where $T_K \approx T_S$, might define the region
where the valence of the Ce-ion starts to increase. 

Band structure calculations \cite{Yamagami94} supported e.~g.~by de
Haas-van Alphen (dHvA) measurements \cite{King91} have shown that the
4f$^1$ Ce electron does not participate to the Fermi surface of \crg
as it is the case for \crs \cite{Onuki96,Aoki2001,Zwicknagl92}. The
comparison between dHvA frequency-branches of \crg and the theoretical
branches of LaRu$_2$Ge$_2$ implies that the 4f electron in \crg is
fully localised in the ground state \cite{Yamagami94}.  One of the
original motivation of the present transport investigation was to
follow the promotion of this 4f electron. However, the data do not
show any distinct pressure or a pressure range where the 4f$^1$
electron becomes delocalized in CeRu$_2$Ge$_2$.  Neither the residual
resistivity \cite{Jaccard99} nor a plot of $S/T$ at 1.5 K reveals a
distinct feature related to the promotion of the 4f$^1$
electron. $S/T$ at 1.5 K has a marked maximum around 6~GPa but it
seems to be correlated to $T_L$. It is also noteworthy that close to
$p_c$, $S/T$ vs $T$ is without any anomaly and does not confirm the
predictions of Ref.~\cite{Paul2001}. Thus, the question can be asked
whether the nature of the pressure-induced quantum discontinuity is of
second order or if it really exists in CeRu$_2$Ge$_2$.

\section{Conclusion}
The influence of pressure on the temperature dependence of the
thermoelectric power $S(T)$ of \crg was measured up to 16~GPa. The
various magnetic phase transitions below 10~K yield a complex $S(T)$
behaviour. A large positive peak in $S(T)$ centered at $T_S$ just
below room temperature starts to develop at pressures above 3~GPa. It
is ascribed to the interplay between incoherent Kondo scattering and
crystal-field effects. A pressure-induced low-temperature maximum in
$S(T)$ at $T_K$ develops in the range 5.7~GPa~$\leq p \leq8$~GPa. It is
interpreted as a signature of the Kondo effect, since its position
shows a similar pressure dependence as $1/\sqrt{A(p)}$. This implies
that only one energy scale seems to exists in
CeRu$_2$Ge$_2$. A revised $(T,p)$ phase diagram, based on transport
and calorimetric investigations on samples of the same batch, suggests
that in \crg long-range magnetic order is suppressed at a
critical pressure $p_c=7.8$~GPa. Well above $p_c$, the $T_K(p)$ and
$T_S(p)$ dependences merge at a pressure $p_v\approx 11$~GPa, defining
a pressure range where the Ce-valence starts to increase.

\begin{acknowledgments} We thank T.~C.~Kobayashi, Osaka University, who
assisted the experiment in its early stage during his stay in Geneva.
The help of M.~Malquarti is acknowledged who performed some of the
measurements as part of his university training. H. W. is grateful to
B. Coqblin, P. Haen, and J. A. Mydosh for stimulating discussions. This work was
supported by the Swiss National Science Foundation.
\end{acknowledgments}


\begin{thebibliography}{8.}
\bibitem{oscillations} The tiny oscillation below 10~K is very likely
 an artefact of the measurement. The origin of the weak anomaly
 ($\approx 1~\mu$V/K) close to 70~K is not evident; a similar
 structure was also observed in LaPd$_2$Si$_2$ \cite{Bando93}.
\end{thebibliography}
\end{document}